# Strain tuning of charge density wave and Mott-insulating states in monolayer $VTe_2$

Wenqian Tu,[1,2] Run Lv,[1,2] Dingfu Shao,[1] Yuping Sun,[3,1,4] and Wenjian Lu[1,a)]

[1]Key Laboratory of Materials Physics, Institute of Solid State Physics, HFIPS, Chinese Academy of Sciences, Hefei 230031, China

[2]University of Science and Technology of China, Hefei 230026, China

[3]High Magnetic Field Laboratory, HFIPS, Chinese Academy of Sciences, Hefei 230031, China

[4]Collaborative Innovation Center of Microstructures, Nanjing University, Nanjing 210093, China

[a)] Author to whom correspondence should be addressed: wjlu@issp.ac.cn

Monolayer vanadium ditelluride ($VTe_2$) exhibits a $2\sqrt{3}\times2\sqrt{3}$ charge density wave (CDW) order intertwined with a Mott-insulating state. However, the physical mechanisms driving the emergence of CDW order and Mott-insulating state are still not well understood. In this study, we systematically investigate the electronic band structure, phonon dispersion, and electron-phonon coupling (EPC) of monolayer $VTe_2$ under applied biaxial strain. Our results reveal that the $2\sqrt{3}\times2\sqrt{3}$ CDW phase is metastable in free-standing monolayer $VTe_2$ and becomes stabilized under compressive strain below $\varepsilon$ = -2%. The formation of CDW order originates dominantly from strong EPC effect, rather than Fermi surface nesting. The narrowing of the bandwidth due to the CDW order, combined with the correlation effect of the V-3*d* orbital, collectively drives the system into a Mott-insulating state. Furthermore, we find that tensile strain suppresses CDW order and induces a superconducting state above a critical strain threshold ($\varepsilon$ = 2%). These findings enhance our understanding of correlation physics in monolayer $VTe_2$ and provide a pathway for strain-engineered manipulation of quantum phases in two-dimensional transition metal dichalcogenides.

## I. INTRODUCTION

Layered transition metal dichalcogenides (TMDs) have attracted significant attention due to their rich physical properties, including charge density wave (CDW) order, Mott physics, and superconductivity (SC) [1-3]. The question of whether CDW and SC are competing or compatible remains a long-standing enigma. For instance, in 1*T*-$TaS_2$, SC emerges within a textured CDW order, suggesting their coexistence [4,5]. Conversely, in both 1*T*-$TiSe_2$ and 2*H*-$TaS_2$, SC is enhanced when CDW is suppressed [6-8]. Among TMDs, only a select few exhibit Mott-insulating properties. A prominent example is provided by 1$T$-$TaS_2$, where the Mott-insulating state coexists with the so-called "star-of-David" CDW order. Various methods to manipulate the CDW order and induce the collapse of the Mott-insulating state in 1$T$-$TaS_2$ have been extensively explored, including high pressure [4,9], doping [10-13], and interlayer stacking [14,15]. The realization of two-dimensional (2D) forms of TMDs has caused the CDW order and the Mott-insulating state to be revisited. For instance, monolayer (ML) 1$T$-$TaSe_2$ has been identified as a strongly correlated 2D Mott insulator [16,17], in contrast to its bulk counterpart, which exhibits metallic properties due to weakened Coulomb interactions from increased electronic delocalization and screening.

Recently, ML $VTe_2$ stabilized in the 1$T$ phase has been found to exhibit a coexistence of metallic $4\times4$ CDW and insulating $2\sqrt{3}\times2\sqrt{3}$ CDW orders in different regions at low temperatures [18-20]. In contrast, bulk $VTe_2$ crystallizes in a monoclinic $1T''$ phase with $3\times1\times3$ periodic lattice distortions [21]. The emergence of the insulating state accompanied by $2\sqrt{3}\times2\sqrt{3}$ CDW order in ML $VTe_2$ has been attributed to strong V-3$d$ electron correlation, which induces a Mott-insulating state [20]. Furthermore, previous experiments have also identified $4\times1$ and $5\times1$ CDW orders in ML $VTe_2$ [19,22-24]. The experimentally observed different CDW orders may be ascribed to their sensitivity to external conditions for samples. Since ML materials are typically prepared on substrates, substrate-induced strain is an inevitable factor, providing opportunities for tuning the CDW state. To date, the physical mechanisms underlying CDW formation in ML $VTe_2$ remain poorly understood, and a comprehensive understanding of the interplay between CDW and the Mott state is still elusive. Investigating how strain can modulate the CDW orders and the Mott-insulating state in ML $VTe_2$ could offer critical insights into controlling correlated quantum states in ML $VTe_2$.

In the present work, we employ first-principles calculations to systematically investigate the electronic band structure, phonon dispersion, and electron-phonon coupling (EPC) in ML $VTe_2$ under biaxial strain. Our comprehensive study reveals that the emergence of diverse CDW orders in ML $VTe_2$ can be primarily attributed to the presence of strong EPC. Furthermore, we demonstrate that strain can significantly modulate the competition between different CDW orders, resulting in distinct electronic ground states. The $4\times4$ CDW phase is the ground state in free-standing samples, while the $2\sqrt{3}\times2\sqrt{3}$ CDW phase becomes more energetically favorable under compressive strain. Our calculations demonstrate that the CDW-induced band narrowing and the strong electron correlation effect of the V-3$d$ orbital lead to the Mott-insulating state. With increasing tensile strain, the CDW order is gradually suppressed, consequently reducing the Mott gap. Interestingly, as the Mott-insulating state is fully collapsed, a superconducting state could be induced, with the superconducting transition temperature $T_c$ reaching 8.5 K at 6% tensile strain. These findings not only deepen our understanding of correlation physics in ML $VTe_2$ but also open new avenues for strain engineering of quantum phenomena in 2D TMD materials.

## II. METHODS

The first-principles calculations based on density functional theory (DFT) were performed using the Vienna Ab Initio Simulation Package (VASP) [25,26]. The exchange-correlation function was approximated using the generalized gradient approximation (GGA) formulated by Perdew, Burke, and Ernzerhof (PBE) [27]. To simulate the ML, a 16 Å vacuum layer was introduced to avoid interactions between neighboring slabs. The Coulomb repulsion of V-3*d* electrons was accounted for using the DFT+U mothed with a Hubbard U parameter of 2 eV, consistent with the previous study [20]. The kinetic energy cutoff was set to 500 eV, and the Brillouin zone (BZ) was sampled using a $\Gamma$-centered grid with a resolution of 0.02 $Å^{-1}$ for self-consistent calculations. Transition state energy differences were calculated using the nudged elastic band (NEB) method [28,29], as implemented in the VTST code [30]. The phonon dispersion spectrum for the 1*T* structure was calculated using the QUANTUM-ESPRESSO packages [31] with the density functional perturbation theory (DFPT) [32]. A $10\times10\times1$ *q*-point grid and a $20\times20\times1$ *k*-point grid were employed to sample the BZ. The wave function and charge density energy cutoffs were set to 100 Ry and 700 Ry, respectively. The biaxial strain was simulated by adjusting the lattice parameters along the *a*-*b* axis. The strain magnitude ε is defined as $(a - a_0)/a_0 \times 100\%$, where positive (negative) values of ε denote biaxial tensile (compressive) strain applied to the system.

## III. RESULTS AND DISCUSSION

At room temperature, ML $VTe_2$ crystallizes in a high-symmetry octahedral 1*T* structure (Fig. 1(a)), where hexagonally arranged V atoms are sandwiched between two layers of Te atoms in an octahedral coordination. Upon cooling to lower temperatures, ML $VTe_2$ exhibits two distinct CDW states: a metallic $4\times4$ CDW order and an insulating $2\sqrt{3}\times2\sqrt{3}$ CDW order [18,20]. The low-symmetry CDW superstructure typically evolves from the high-symmetry phase, with distortions introduced by underlying instability. When the temperature drops below the CDW transition temperature, the material undergoes lattice distortion, leading to atoms rearrangement and cluster formation. To investigate the characteristics of ML $VTe_2$, we simulated the superstructures of the $4\times4$ and $2\sqrt{3}\times2\sqrt{3}$ CDW orders. In the $4\times4$ superstructure, 13 V atoms are arranged in a "star of David" cluster (Fig.

1(c)). Conversely, in the $2\sqrt{3}\times2\sqrt{3}$ superstructure, 12 V atoms are arranged into a truncated triangle-shaped cluster (Fig. 1(d)). The lattice distortion is attributed to the dimerization of V atoms accompanied by the vertical bulge of the Te atoms.

FIG. 1. (a) Crystal structure of bulk 1*T*-$VTe_2$. The red (brown) balls represent V (Te) atoms. (b) BZs

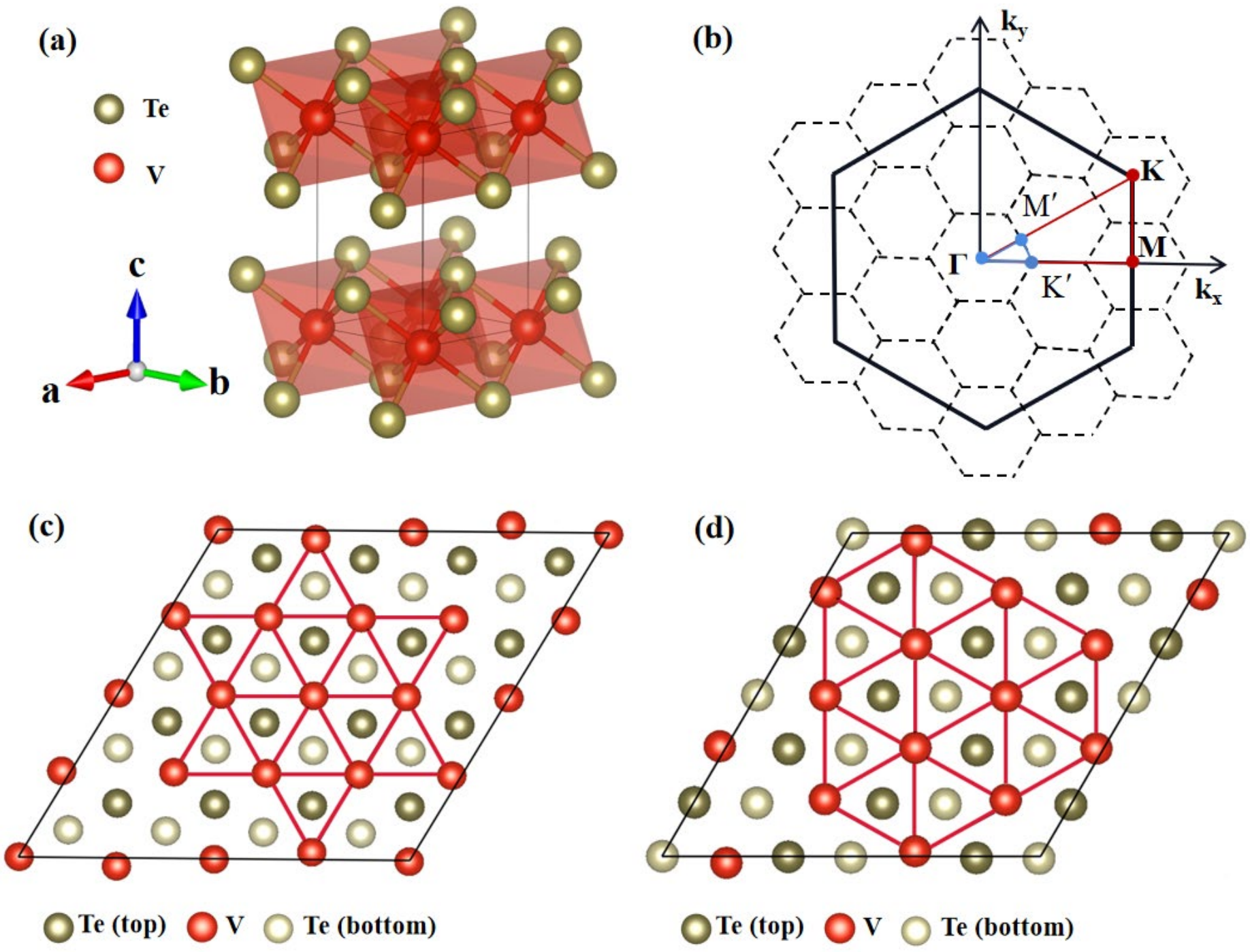

of monolayer 1*T*-$VTe_2$ unit cell (solid line) and $2\sqrt{3}\times2\sqrt{3}$ superstructure (dashed line), in which the high-symmetry *k*-points are labeled by red and blue points, respectively. Superstructures of the (c) $4\times4$ and (d) $2\sqrt{3}\times2\sqrt{3}$ CDW orders, the red colored lines represent the atomic bonds whose lengths change due to the CDW distortions. Here, the V-V bond lengths are decreased by ~ 6% than that of the high-symmetry 1*T* phase.

Our calculations show that the structural optimization of the $2\sqrt{3}\times2\sqrt{3}$ superstructure, incorporating a Hubbard U value of 2 eV, revealed a ~ 6% reduction in the bond length between neighboring V atoms compared to the ideal 1*T* structure. The result is consistent with previous scanning tunneling spectroscopy experimental observations [20]. We calculated the electronic band structures of prototype 1*T*, $4\times4$ CDW, and $2\sqrt{3}\times2\sqrt{3}$ CDW phases of ML $VTe_2$. As shown in Figs. 2 (a)-(c), both the prototype 1*T* and the $4\times4$ CDW phases exhibit metallic behavior, while a full gap opens for the $2\sqrt{3}\times2\sqrt{3}$ CDW phase. The electron bands near the Fermi level are contributed mainly by the V-3*d* orbitals. For the $2\sqrt{3}\times2\sqrt{3}$ CDW phase, the CDW order narrows the bandwidth, allowing the strong correlation effect of the V-3*d* orbital to open a Mott gap and induce an insulating state, in agreement with previous experiment results [18,20]. The Mott physics in the $2\sqrt{3}\times2\sqrt{3}$ CDW phase of ML $VTe_2$ is different from other widely explored TMDs, such as ML 1*T*-$TaS_2$, 1*T*-$TaSe_2$, and 1*T*-$NbSe_2$ [16,33-36]. In those materials, the "star of David" cluster hosts an odd number (13) of *d*-electrons per CDW superstructure and an isolated half-filled narrow band. In contrast, ML $VTe_2$ nominally hosts an even number (12) of *d*-electrons. The distinction suggests that the above-mentioned Mott-insulating model may not directly apply to $VTe_2$.

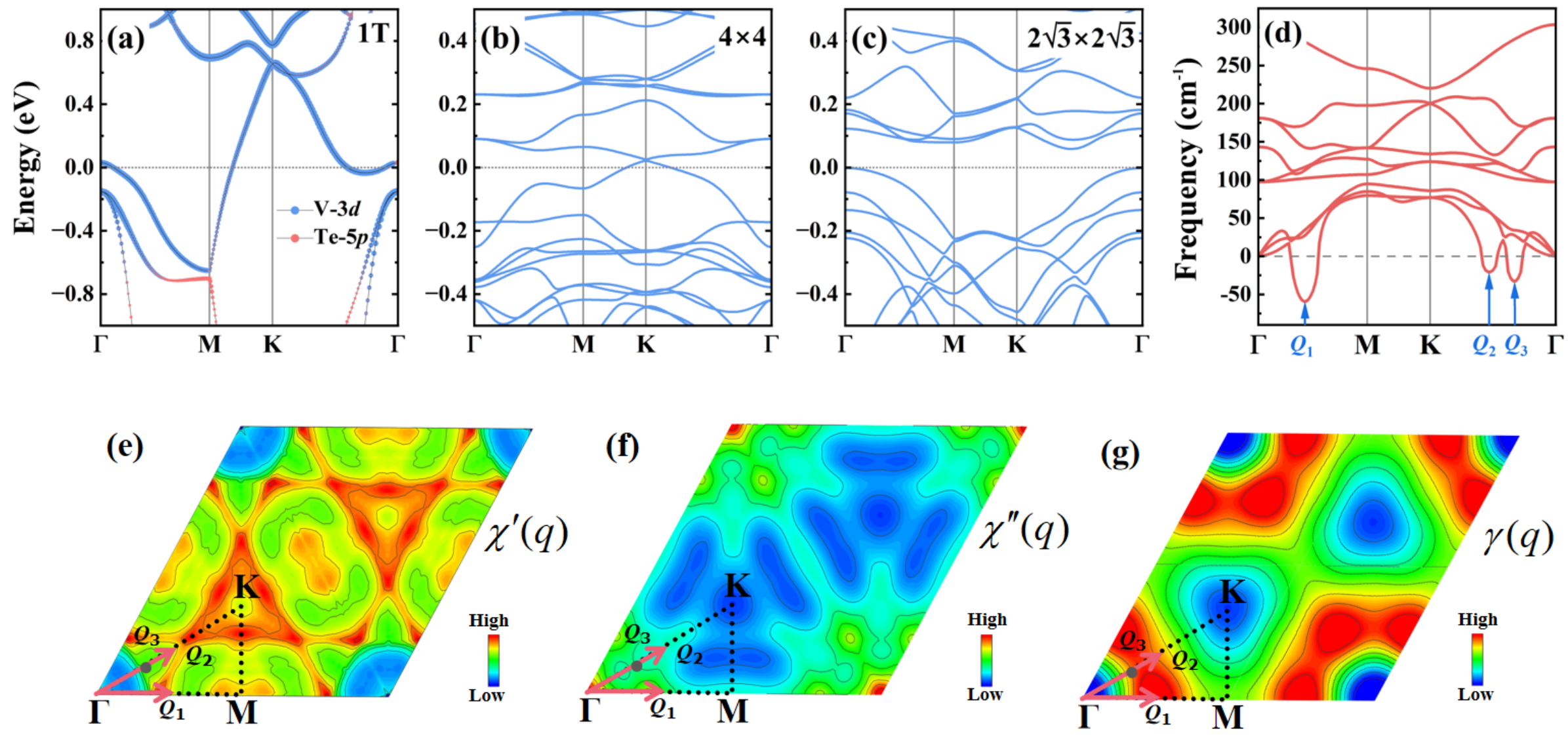

FIG. 2. Calculated band structures of ML $VTe_2$ in the (a) prototype 1*T* (b) 4×4 CDW and (c) $2\sqrt{3}\times2\sqrt{3}$ CDW phases using DFT + U (U = 2 eV) method. (d) Phonon dispersion spectrum of ML $VTe_2$ with the high-symmetry 1*T* structure, where $Q_1$, $Q_2$, and $Q_3$ stand for three distinct soft phonon modes (indicated by blue arrows). (e) Real and (f) imaginary parts of the electron susceptibility cross-section of ML $VTe_2$ in the plane of $q_z = 0$, where $Q_1$ and $Q_2$ are indicated by pink arrows, and $Q_3$ is labeled by a gray point. (g) Phonon linewidth of the lowest phonon mode of ML $VTe_2$ in the $q_z = 0$ plane.

To investigate the physical origin of CDW in ML $VTe_2$, we first calculated the phonon spectrum of its high-temperature 1*T* structure. For materials exhibiting CDW orders at low temperatures, the phonon spectrum of the structure with a high-temperature phase typically displays instability near the CDW wave-vector $Q_{CDW}$ [37]. Accordingly, in real space, the atom displacements form the low-temperature CDW superstructure. In Table S1 of Supplemental Material, we list the $Q_{CDW}$ of various CDW superstructures. The calculated phonon spectrum in Fig. 2(d) shows three distinct soft phonon modes labeled $Q_1$, $Q_2$, and $Q_3$. The phonon instability at the wave vector $Q_1 \sim \frac{1}{2}\Gamma M$ is related to the $4\times4$ CDW order, while the one at the wave vector $Q_2 \sim \frac{1}{2}\Gamma K$ associates to the $2\sqrt{3}\times2\sqrt{3}$ CDW order. With respect to the wave vector $Q_3 \sim \frac{1}{3}\Gamma K$, we attribute it to the $\sqrt{19}\times\sqrt{19}$ CDW order (supplemental Fig. S1), previously experimentally observed in 1$T$-$NbTe_2$ and 1$T$-$TaTe_2$ [38-40], which are structurally analogous to 1$T$-$VTe_2$.

Given that the EPC plays an important role in CDW formation [41,42], we calculate the phonon linewidth $\gamma$, directly tied to the strength of the EPC. The wave-vector $q$ and phonon mode $\nu$ decomposed $\gamma_{q\nu}$ is defined as

$$\gamma_{\mathrm{q}\nu} = 2\pi\omega_{q\nu}\sum_{ij}\int\frac{d^3k}{\Omega_{BZ}}\left|g_{q\nu}(k,i,j)\right|^2 \times \delta(\varepsilon_{q,i}-\varepsilon_F)\delta(\varepsilon_{k+q,i}-\varepsilon_F). \quad (1)$$

Here $g_{q\nu}(k,i,j)$ is the EPC coefficient and can be calculated by

$$g_{q\nu}(k,i,j) = (\frac{\hbar}{2M\omega_{q\nu}})^{1/2}\left\langle\psi_{i,k}\left|\frac{dV_{SCF}}{d\hat{u}_{q\nu}}\hat{\xi}_{q\nu}\right|\psi_{j,k+q}\right\rangle, \quad (2)$$

where $\psi$ is the wave function, $V_{SCF}$ is the Kohn-Sham potential, $\hat{u}$ is atomic displacement, and $\hat{\xi}$ is the phonon eigenvector. The calculation results in Fig. 2(g) show that the maximum of $\gamma$ aligns with wave vectors $Q_1$ and $Q_2$, while away from wave vector $Q_3$. This suggests that the EPC interaction plays an important role in the formation of the $4\times4$ and $2\sqrt{3}\times2\sqrt{3}$ CDW orders in ML $VTe_2$.

Another perspective of the driving force of CDW formation is the Fermi surface (FS) nesting, which can be evaluated quantitatively by calculating electron susceptibility [40,43,44]. The real part of the electron susceptibility $\chi'$ is defined as

$$\chi'(q) = \sum_k\frac{f(\varepsilon_k)-f(\varepsilon_{k+q})}{\varepsilon_k-\varepsilon_{k+q}}, \quad (3)$$

where $f(\varepsilon_k)$ is the Fermi-Dirac distribution function. The imaginary part of the electron susceptibility $\chi''$ can be calculated by

$$\chi''(q) = \sum_k\delta(\varepsilon_k-\varepsilon_F)\delta(\varepsilon_{k+q}-\varepsilon_F). \quad (4)$$

$\chi'$ defines the stability of the electronic subsystem, while $\chi''$ is referred to as the "nesting function". If CDW order arises from FS nesting, both $\chi'$ and $\chi''$ should exhibit maximum values at $Q_{\mathrm{CDW}}$. However, our calculations for ML $VTe_2$, as shown in Figs. 2(e) and (f), reveal that neither $\chi'$ nor $\chi''$ displays maxima at the wave vectors $Q_1$, $Q_2$ or $Q_3$. This observation suggests that FS nesting does not play an important role in the formation of the CDW orders in ML $VTe_2$. Our calculations of the accurate phonon linewidth $\gamma$ obtained by DFPT and the electron susceptibility reveal that EPC rather than FS nesting governs the CDW formation in ML $VTe_2$.

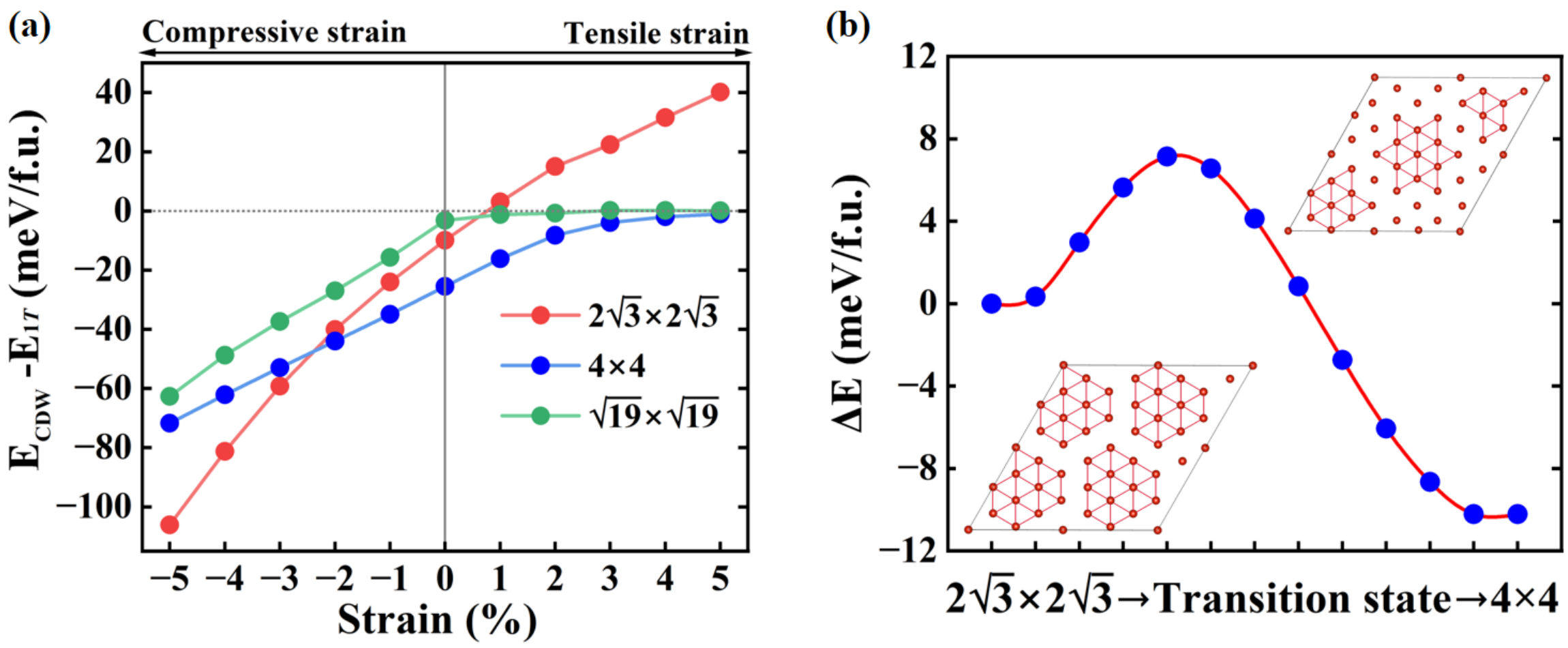

FIG. 3. (a) CDW formation energy of the potential CDW superstructures in ML $VTe_2$ as a function of biaxial strain. (b) Energy difference of the transition from the initial $2\sqrt{3}\times2\sqrt{3}$ CDW phase to the final $4\times4$ CDW phase. The insets show the schematic illustrations of the $2\sqrt{3}\times2\sqrt{3}$ and $4\times4$ superstructures.

To further evaluate the stability of potential CDW phases from an energy perspective, we investigate how biaxial strains affect the formation energy of the $4\times4$, $2\sqrt{3}\times2\sqrt{3}$, and $\sqrt{19}\times\sqrt{19}$ superstructures. The CDW formation energy is defined as the total energies of the CDW phases relative to that of the high-symmetry $1T$ phase ($E_{CDW}-E_{1T}$). A negative CDW formation energy indicates that the CDW structure is energetically favorable, with more negative values corresponding to greater stability. As shown in Fig. 3(a), under compressive strain, the CDW formation energy becomes negative and its absolute value progressively increases, indicating enhanced stability of the CDW phases. In contrast, under tensile strain, the formation energy approaches zero or even becomes positive, suggesting that the CDW order is suppressed. This conclusion is further supported by analyzing the displacement of V atoms in potential CDW phases under different biaxial strains, as shown in supplemental Fig. S2. The distortion of V atoms decreases under tensile strain, which correlates with an inhibitory effect on CDW orders. Collectively, these findings demonstrate that compressive strain promotes CDW formation in ML $VTe_2$, while tensile strain hinders it.

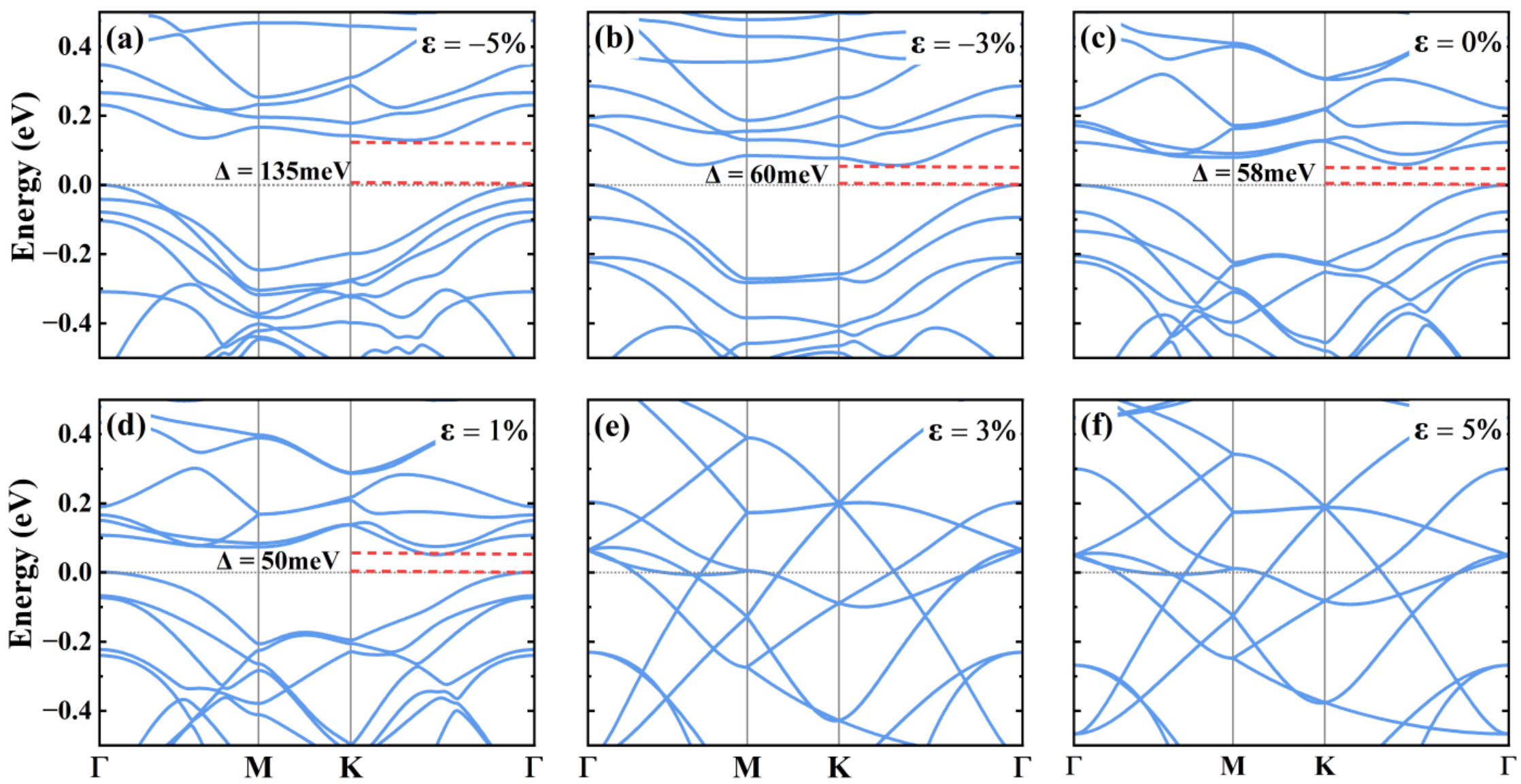

FIG. 4. Electronic band structures of the $2\sqrt{3}\times2\sqrt{3}$ CDW phase under biaxial strain: (a, b) for compressive strain, (c) without strain, and (d)-(f) for tensile strain. The top of the valence band is set to zero. Red dashed lines in (a)-(d) indicate the energy gap Δ, which exhibits a systematic reduction as the magnitude of biaxial tensile strain increases.

In free-standing ML $VTe_2$ without strain, the $4\times4$ CDW phase exhibits the largest absolute formation energy, while that of the $\sqrt{19}\times\sqrt{19}$ CDW phase shows the smallest. This implies that the $4\times4$ CDW phase is the most stable in free-standing ML $VTe_2$ and the $\sqrt{19}\times\sqrt{19}$ CDW phase may be unstable. The absence of experimental observations of the $\sqrt{19}\times\sqrt{19}$ CDW phase until now could be attributed to this instability. Furthermore, the $2\sqrt{3}\times2\sqrt{3}$ superstructure likely represents a metastable phase. However, under compressive strain exceeding 2%, a notable transition occurs: The absolute formation energy of the $2\sqrt{3}\times2\sqrt{3}$ CDW phase surpasses that of the $4\times4$ CDW phase, indicating a shift in ground state stability. This suggests that experimentally observed $2\sqrt{3}\times2\sqrt{3}$ CDW phase in strained ML $VTe_2$ samples could be attributed to such compressive strain effects. To further explore the transition dynamics between the CDW phases, we employed the NEB method to evaluate the energy barrier from the initial $2\sqrt{3}\times2\sqrt{3}$ CDW phase to the final $4\times4$ CDW phase. The energy difference is defined as $\Delta E = E_{2\sqrt{3}\times2\sqrt{3}} - E_{transition}$. As shown in Fig. 3(b), though the $4\times4$ CDW phase is energetically favorable, the system may remain locked in the metastable $2\sqrt{3}\times2\sqrt{3}$ CDW phase in free-standing ML $VTe_2$ without strain due to the high activation barrier (~ 8meV). Transition to the more stable $4\times4$ CDW phase requires sufficient energy to overcome the hill in the free energy space and stabilize the system in a new local valley.

Further analysis of the electronic states under strain reveals significant differences. For the $2\sqrt{3}\times2\sqrt{3}$ CDW phase, the Mott gap increases upon compressive strain (Figs. 4(a-c)). We found that applying compressive strain reduces the lattice parameters and enhances the atomic displacement, thereby enhancing the CDW distortion. Consequently, the bandwidth narrows and the Mott gap gradually widens. Conversely, tensile strain suppresses the CDW distortion, leading to a reduction in the Mott gap. Notably, when tensile strain exceeds 3%, the energy gap disappears completely (Fig. 4(e)), indicating the collapse of the Mott-insulating state. In contrast to the $2\sqrt{3}\times2\sqrt{3}$ CDW phase, the metal-characteristic energy band of the $4\times4$ CDW phase remains relatively insensitive to strain variation (supplemental Fig. S3), demonstrating its robustness against external mechanical perturbations.

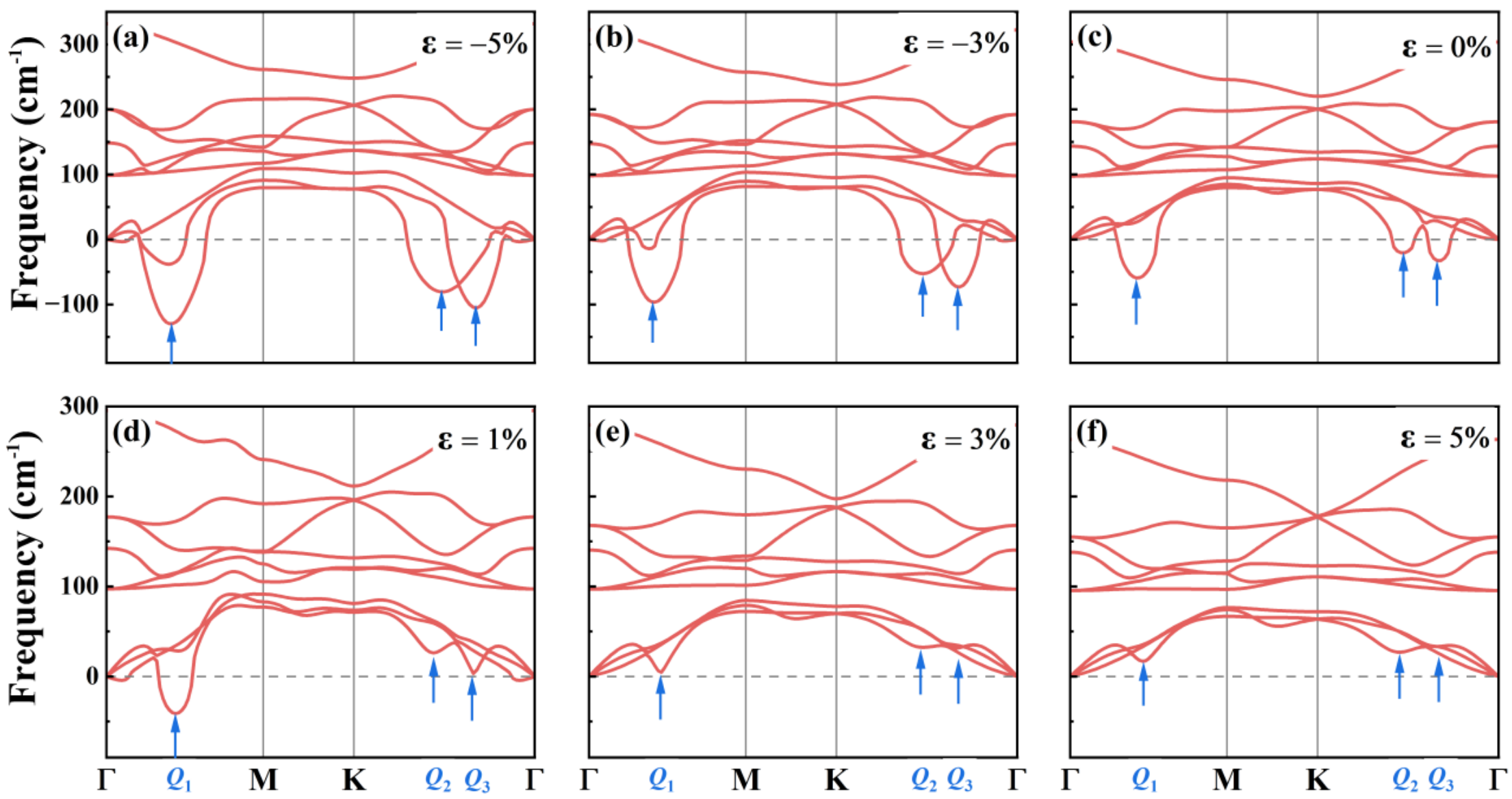

FIG. 5. Phonon dispersion spectra of ML $VTe_2$ in the high-symmetry $1T$ phase under biaxial strain: (a) and (b) for compressive strain, (c) without strain, and (d)-(f) for tensile strain, where $Q_1$, $Q_2$, and $Q_3$ stand for three distinct soft phonon modes (indicated by blue arrows).

Next, we calculate the phonon spectrum to evaluate the stability of the high-symmetry $1T$ structure under various biaxial strains. It is well-established that imaginary phonon frequencies at a specific wave vector $Q_{\mathrm{CDW}}$ indicate lattice restructuring associated with a CDW superstructure characterized by $Q_{\mathrm{CDW}}$. Conversely, the absence of imaginary frequency in the phonon spectrum implies structural stability relative to CDW superstructure. Figure 5 shows the phonon dispersion spectra of ML $VTe_2$ with undistorted 1$T$ structure under different strains. Compared to the strain-free case (Fig. 5(c)), the acoustic branches at wave vectors $Q_1$, $Q_2$, and $Q_3$ exhibit increased softening and become less stable as compressive strain increases. Under tensile strain, however, these softened acoustic branches become more stabilized, suggesting that CDW distortions are suppressed with increasing tensile strain. Interestingly, our findings align with those observations in 1$H$-$TaSe_2$ [45], yet they present a contrasting scenario to previous studies on 1$T$-$TaS_2$ [9] and 1$T$-$TiSe_2$ [46]. This divergence underscores the unique mechanical and electronic properties of ML $VTe_2$ under strain.

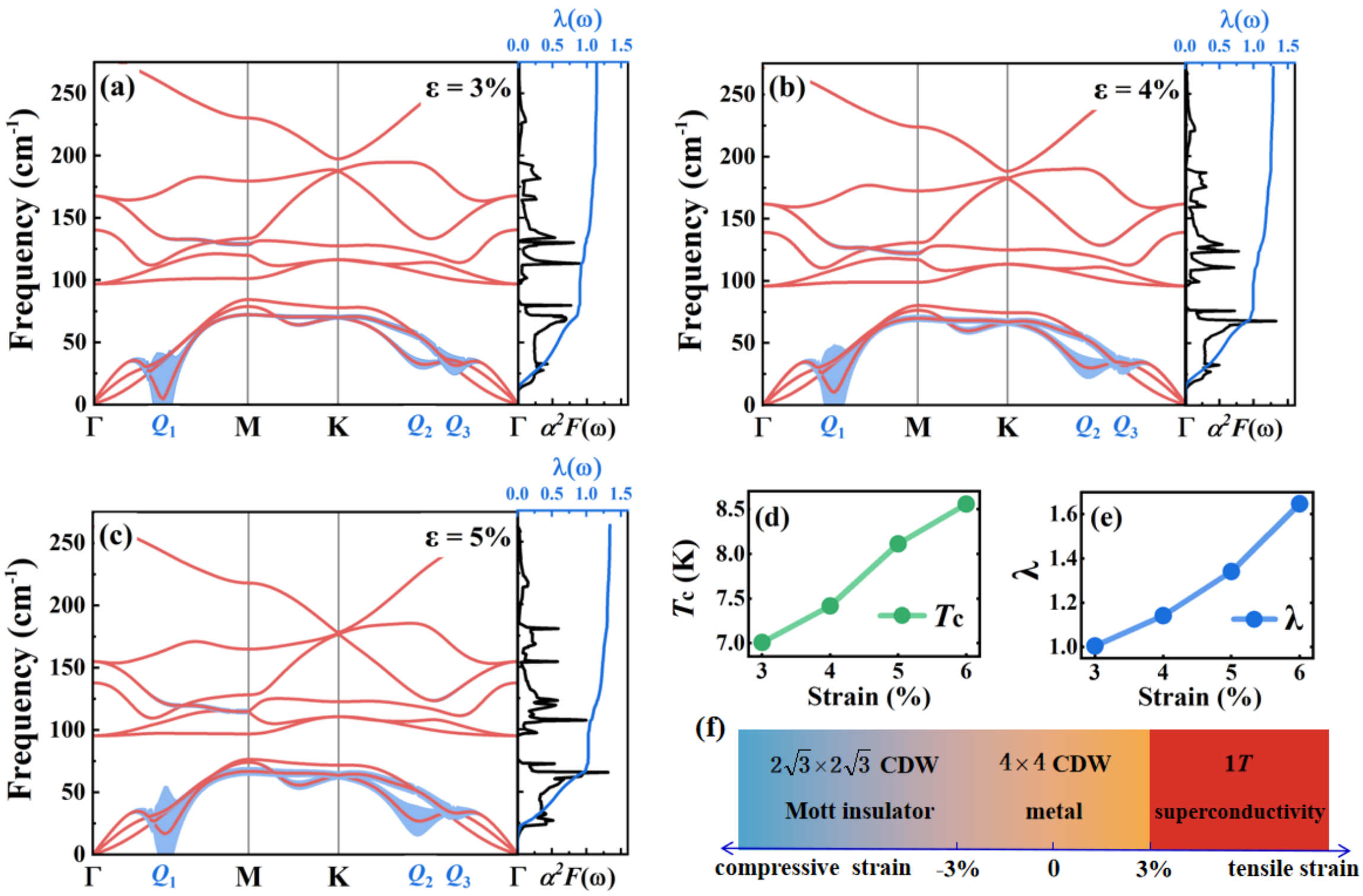

FIG. 6. Phonon dispersion spectra weighted by the magnitude of EPC strength $\lambda_{qv}$ under (a) 3%, (b) 4%, and (c) 5% biaxial tensile strain. The Eliashberg spectral function $\alpha^2F(\omega)$ (black) and the integrated strength of EPC constant $\lambda(\omega)$ (blue) are plotted on the right side. (d, e) Calculated superconducting transition temperature $T_c$ and EPC constant $\lambda(\omega)$ as a function of biaxial tensile strain (3%-6%). (f) Schematic diagram illustrating the structural and electronic state changes in ML $VTe_2$ under varying biaxial strain.

SC typically emerges after the suppression of CDW or Mott-insulating states, as observed in most TMDs [4,47,48]. Here, we investigate the possibility of a superconducting state in ML $VTe_2$ under large tensile strain, where both CDW and Mott-insulating states are completely suppressed. To evaluate SC, we calculated the Eliashberg spectral function $\alpha^2F(\omega)$ and EPC strength $\lambda(\omega)$ using the

following expressions:

$$\alpha^2F(\omega) = \frac{1}{2\pi N(E_F)}\sum_{qv}\delta(\omega-\omega_{qv})\frac{\gamma_{qv}}{\hbar\omega_{qv}}, \tag{5}$$

and

$$\lambda(\omega) = \sum_{qv}\lambda_{qv} = 2\int\frac{\alpha^2F(\omega)}{\omega}d\omega. \tag{6}$$

The calculated $\alpha^2F(\omega)$ and integrated $\lambda(\omega)$ are plotted on the right side of phonon spectra in Figs. 6(a)-(c). Additionally, we decorate the calculated EPC strength at each wave vector and phonon-mode $\lambda_{qv}$. Notably, the phonon branch with the CDW-type softening mode exhibits the largest electron-phonon coupling. Next, we calculate superconducting transition temperature $T_c$ using the Allen-Dynes-modified McMillan formula [49]:

$$T_c = \frac{\omega_{\log}}{1.2}\exp\left(-\frac{1.04(1+\lambda)}{\lambda-\mu^*-0.62\lambda\mu^*}\right), \tag{7}$$

where the Coulomb pseudopotential $\mu^*$ is set to a typical value of 0.1 [42,50]. We estimated the $T_c$ and $\lambda$ of ML $VTe_2$ under varying tensile strain and show the results in Figs. 6(d) and (e). Both $T_c$ and $\lambda$ exhibit a monotonous increase with increasing tensile strain. Specifically, at ε = 3%, we determine $T_c$ = 7 K and $\lambda$ = 1.1. These values reach their maximum at ε = 6% with $T_c$ = 8.5 K and $\lambda$ = 1.6. Our results indicate that tensile stress suppresses the CDW order and Mott gap in the $2\sqrt{3}\times2\sqrt{3}$ CDW state, while simultaneously enhancing superconducting properties. In contrast, compressive strain notably strengthens the CDW state. These results demonstrate that biaxial strain provides an effective method of manipulating the CDW, Mott-insulating, and superconducting states in ML $VTe_2$, as summarized schematically in Fig. 6(f). This strain-dependent behavior underscores the potential of mechanical deformation as a flexible tool for controlling the electronic properties of ML $VTe_2$ with correlated electron states.

## IV. CONCLUSION

In conclusion, we have systematically investigated the physical origins of CDW orders and the effects of biaxial strain on CDW phases and electronic states in ML $VTe_2$. Our results reveal the critical role of EPC in driving the formation of CDW in ML $VTe_2$. The $4\times4$ CDW phase represents the ground state in free-standing ML $VTe_2$, while the $2\sqrt{3}\times2\sqrt{3}$ CDW phase becomes more energetically favorable under compressive strain. The narrowing of bandwidth due to the CDW order, combined with the strong correlation effect of the V-3*d* orbital, results in the opening of a Mott gap in the $2\sqrt{3}\times2\sqrt{3}$ CDW phase. Our calculations of the band structure and phonon spectra under various biaxial strains indicate the profound effects of strain on the electronic properties. Compressive strain significantly enhances the CDW order and the associated Mott-insulating state, while tensile strain suppresses the CDW order, thereby facilitating the emergence of potential SC. Our work highlights the potential of strain engineering as a powerful tool for manipulating phase transitions in ML $VTe_2$. The coexistence of the CDW, Mott-insulating states, and strain-induced superconducting state in ML $VTe_2$ presents promising opportunities for exploring the intricate interplay between distinct quantum states.

## ACKNOWLEDGEMENTS

This work was supported by the National Key Research and Development Program of China under Contract No. 2022YFA1403203. The calculations were performed at Hefei Advanced Computing Center.

SUPPLEMENTAL MATERIAL

# Strain manipulation of charge density wave and Mott-insulating states in monolayer $VTe_2$

Wenqian Tu,[1,2] Run Lv,[1,2] Dingfu Shao,[1] Yuping Sun,[3,1,4] and Wenjian Lu[1,a)]

[1]Key Laboratory of Materials Physics, Institute of Solid State Physics, HFIPS, Chinese Academy of Sciences, Hefei 230031, China

[2]University of Science and Technology of China, Hefei 230026, China

[3]High Magnetic Field Laboratory, HFIPS, Chinese Academy of Sciences, Hefei 230031, China

[4]Collaborative Innovation Center of Microstructures, Nanjing University, Nanjing 210093, China

[*]Corresponding author: wjlu@issp.ac.cn

## 1. Schematic diagram of ML $VTe_2$ in the $\sqrt{19} \times \sqrt{19}$ CDW phase

In the $\sqrt{19}\times\sqrt{19}$ superstructure, both the seven-atom cluster and the triangular cluster of V atoms emerge. The lattice distortion is attributed to the dimerization of V atoms accompanied by the vertical bulge of the Te atoms.

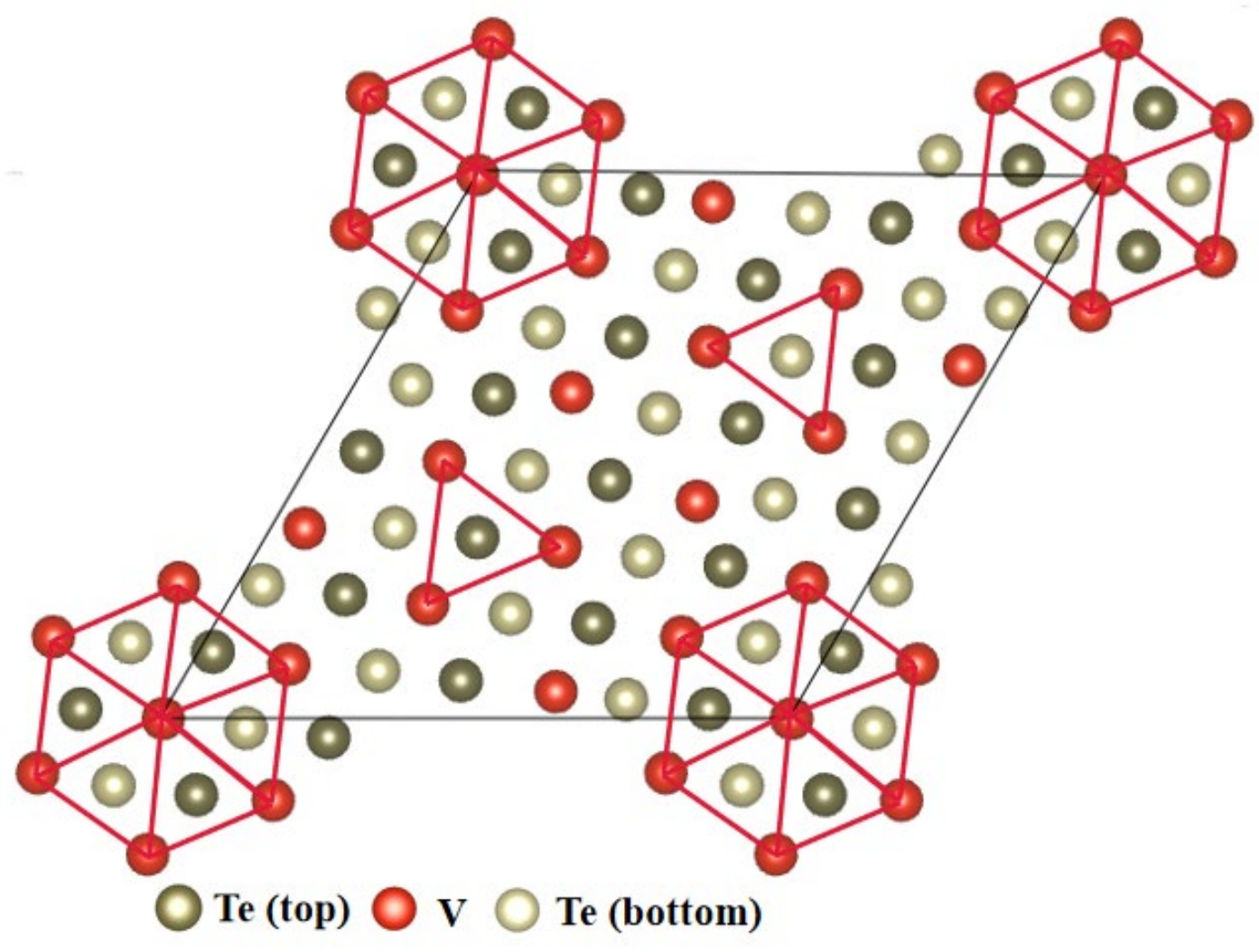

FIG. S1. Superstructure of $\sqrt{19}\times\sqrt{19}$ CDW phase. The red colored lines represent the atomic bonds, whose lengths change due to the CDW distortion. Here, the V-V bond lengths are decreased by ~ 6% than that of the high-symmetry 1$T$ phase.

**2. Average displacement of V atoms in the 4×4 and $2\sqrt{3}\times2\sqrt{3}$ CDW phases as a function of biaxial strain**

Here we investigate the influence of biaxial strain on the 4×4 and $2\sqrt{3}\times2\sqrt{3}$ superstructures. We calculate the displacement of V atoms from high symmetry positions, which can be defined as $D_V=\frac{R_0-R}{R_0}\times100\%$. Here, $R_0$ and $R$ denote the distance of neighbor V atoms before and after lattice distortion associated with CDW order. As shown in Fig. S2, when the compressive strain is applied to the ML $VTe_2$, the displacement of V atoms in both 4×4 and $2\sqrt{3}\times2\sqrt{3}$ CDW phases increases gradually, suggesting that the compressive strain can enhance the CDW orders. On the contrary, when the tensile strain is applied, the displacement of V atoms decreases gradually, which indicates that the tensile strain has an inhibitory effect on the CDW orders.

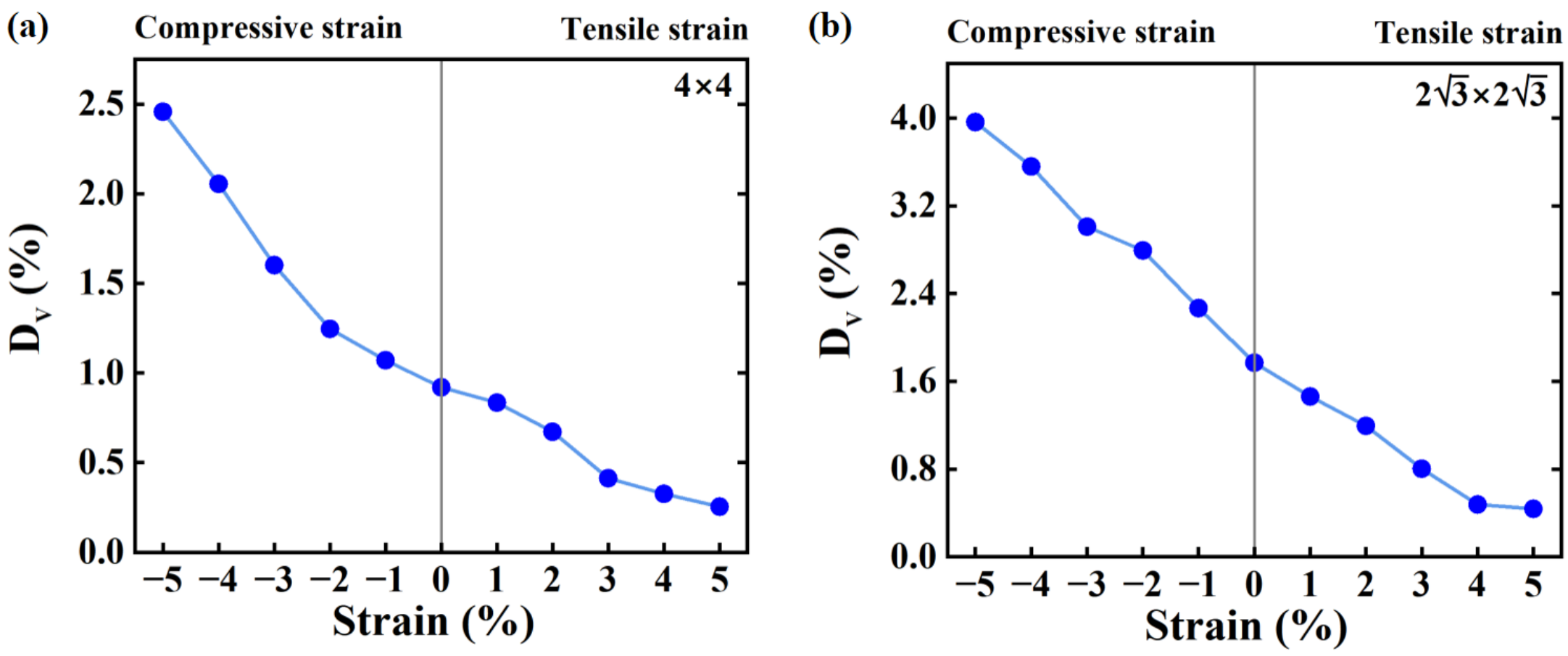

FIG. S2. Displacement of V atoms from high symmetry positions in the (a) 4×4 and (b) $2\sqrt{3}\times2\sqrt{3}$ CDW phases as a function of biaxial strain.

**3. Influence of biaxial strain on the electronic band structures of ML $VTe_2$ in the 4×4 CDW phase**

We investigated the influence of biaxial strain on the electronic band structures of ML $VTe_2$ in the 4×4 CDW phase. As shown in Fig. S3, the metal-characteristic energy bands of the 4×4 CDW phase are insensitive to strain variation.

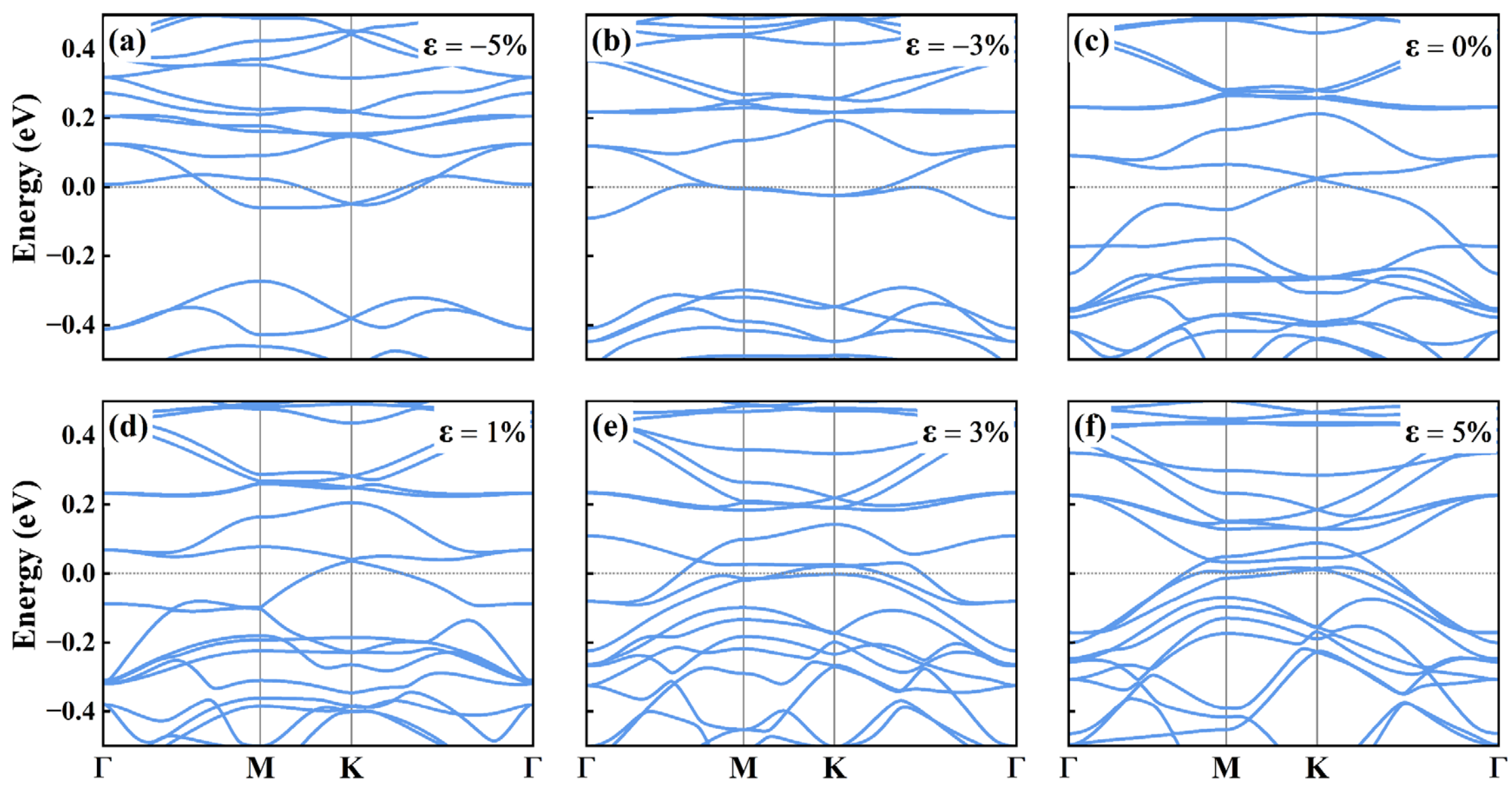

FIG. S3. Electronic band structures of ML $VTe_2$ in the $4\times4$ CDW phase under biaxial strain: (a) and (b) for compressive strain, (c) without strain, and (d)-(f) for tensile strain.

## 4. Lattice vectors of potential CDW superstructures of ML $VTe_2$ in real and reciprocal space

We simulate the potential CDW superstructures in ML $VTe_2$ to find the ground state and further clarify the origin of the multiple CDW order formation. We select the lattice vectors of potential CDW orders as listed in Table S1, which are commensurate with the high-symmetry $1T$ structure. Then, we set the ~ 3% displacement of V atoms while considering the related symmetry to construct the CDW superstructures.

TABLE S1. Lattice vectors of diverse CDW superstructures in real and reciprocal space as a function of the lattice vectors of the high symmetry structure. Here, $a$ and $b$ ($a^*$ and $b^*$) are the basis vectors of the high-symmetry $1T$ structure in real (reciprocal) space, while $a'$ and $b'$ ($q_1$ and $q_2$) are the lattice vectors of CDW superstructures in real (reciprocal) space.

| | $4\times4$ | $2\sqrt{3}\times2\sqrt{3}$ | $\sqrt{19}\times\sqrt{19}$ |
|---|---|---|---|
| $a'$ | $4a$ | $-4a-2b$ | $2a+5b$ |
| $b'$ | $4b$ | $2a+4b$ | $5a+3b$ |
| $q_1$ | $\frac{1}{4}a^*$ | $\frac{1}{3}a^*-\frac{1}{6}b^*$ | $\frac{3}{19}a^*-\frac{5}{19}b^*$ |
| $q_2$ | $\frac{1}{4}b^*$ | $-\frac{1}{6}a^*+\frac{1}{3}b^*$ | $-\frac{5}{19}a^*+\frac{2}{19}b^*$ |